\documentclass[submitting]{nst}
\usepackage{tikz}
\usepackage{graphicx}
\usetikzlibrary{decorations.pathmorphing}
\usepackage{subfigure,dcolumn}
\usepackage[T2A,T1]{fontenc}
\usepackage[russian,english]{babel}
\usepackage{comment}
\usepackage{slashed}
\usepackage{mathtools}
\usepackage{listings}

\begin{document}

\title{Centrality Manipulation in Exclusive Photoproduction at the Electron-Ion Collider}\thanks{Supported in part by the National Key Research and Development Program of China under Contract No. 2022YFA1604900 and the National Natural Science Foundation of China (NSFC) under Contract No. 12175223 and 12005220. W. Zha is supported by Anhui Provincial Natural Science Foundation No. 2208085J23 and Youth Innovation Promotion Association of Chinese Academy of Science.}

\author{Xin Wu}
\affiliation{State Key Laboratory of Particle Detection and Electronics, University of Science and Technology of China, Hefei 230026, China}
\author{Xinbai Li}
\affiliation{State Key Laboratory of Particle Detection and Electronics, University of Science and Technology of China, Hefei 230026, China}
\author{Zebo Tang}
\affiliation{State Key Laboratory of Particle Detection and Electronics, University of Science and Technology of China, Hefei 230026, China}
\author{Kaiyang Wang}
\affiliation{State Key Laboratory of Particle Detection and Electronics, University of Science and Technology of China, Hefei 230026, China}
\author{Wangmei Zha}
\email[Corresponding author, ]{Wangmei Zha Address: No. 96 Jinzhai Road, Hefei city, China Tel: +86 551 63607940  Email: first@ustc.edu.cn}
\affiliation{State Key Laboratory of Particle Detection and Electronics, University of Science and Technology of China, Hefei 230026, China}

\begin{abstract}
In the context of future electron-ion collision experiments, particularly the Electron-Ion Collider (EIC) and the Electron-Ion Collider in China (EicC), investigating exclusive photoproduction processes is of paramount importance. These processes offer a unique opportunity to probe the gluon structure of nuclei across a broad range of Bjorken-$x$, facilitating measurements of nuclear shadowing and searches for gluon saturation and/or the color glass condensate. This paper explores the potential of utilizing neutron tagging from Coulomb excitation of nuclei to effectively determine centrality for exclusive photoproduction in electron-ion collisions. By developing the equivalent photon approximation for fast-moving electrons, this work incorporates a coordinate-space-dependent photon flux distribution, elucidating the relationship between the photon transverse momentum distribution and collision impact parameter. Leveraging spatial information from the photon flux, the differential cross section for Coulomb excitation of nuclei is derived. Our calculations demonstrate that neutron tagging can significantly alter impact parameter distributions, thereby providing a robust method for centrality manipulation in electron-ion collisions. This study contributes essential baseline and strategies for exploring the impact parameter dependence of exclusive photoproduction, offering novel insights for experimental design and data analysis. Ultimately, it provides additional information to better visualize the gluon distribution within the nucleus.
\end{abstract}


\maketitle

\section{Introduction}

Electron-ion collisions present an unparalleled opportunity to explore the internal structure of nucleons and nuclei, particularly the distribution of gluons across different momentum scales. Upcoming facilities, including the Electron-Ion Collider (EIC)~\cite{Accardi:2012qut} in the United States and the Electron-Ion Collider in China (EicC)~\cite{Anderle:2021wcy}, are specifically designed to probe these structures over a wide range of photon virtuality ($Q^2$) and Bjorken-$x$, enabling the study of phenomena such as nuclear shadowing and gluon saturation. The use of high-energy electron beams interacting with protons and heavy ions allows for precise measurements of the spatial and momentum distributions of gluons within the target~\cite{Iancu:2003xm}, which is crucial for understanding quantum chromodynamics (QCD) in dense nuclear environments.

Exclusive photoproduction is one of the key processes for probing gluon distributions within nuclei. In this process, a virtual photon emitted by the electron interacts coherently with the target, leading to the production of a vector meson while leaving the target intact. This process provides a direct probe of the gluon density, as the cross-section is sensitive to the gluon distribution within the target. Specifically, in coherent photoproduction, the photon fluctuates into a quark-antiquark pair, which subsequently scatters elastically from the target through the exchange of a color-neutral object, typically a Pomeron at high energies~\cite{Bartels:2003yj}. Such studies are instrumental in understanding phenomena like gluon shadowing, where gluon densities are suppressed in nuclei compared to free protons, and in providing evidence for gluon saturation and the formation of the color glass condensate~\cite{Jones:2013pga,Ivanov:2004vd,Wang:2023thy,CMS:2023snh}.

To gain insights into the spatial distribution and fluctuations of gluons, measurements of the differential cross-section $\mathrm{d}\sigma/\mathrm{d}t$ are of paramount importance~\cite{Toll:2012mb}. The momentum transfer $t$ is directly related to the transverse distance between the interacting particles, thus providing information on the spatial distribution of gluons. In relativistic heavy-ion collisions, significant progress has been made in probing this distribution through $\mathrm{d}\sigma/\mathrm{d}t$ measurements~\cite{Klein:2019qfb}. Early studies by the STAR experiment at the Relativistic Heavy Ion Collider (RHIC) utilized $\rho$ meson photoproduction to reconstruct the spatial distribution of gluons via an inverse Fourier transform of the $\mathrm{d}\sigma/\mathrm{d}t$ distribution~\cite{STAR:2017enh}. Further advancements were made by the ALICE experiment at the Large Hadron Collider (LHC), which measured $\mathrm{d}\sigma/\mathrm{d}t$ while accounting for the transverse momentum of the photons~\cite{ALICE:2021tyx}, introducing interference effects that allowed for a more detailed analysis of gluon spatial distributions. More recently, STAR~\cite{STAR:2022wfe} reported measurements of $\rho$ meson photoproduction that exploited the linear polarization of photons, adding an extra dimension to the analysis and thereby enhancing sensitivity to spatial anisotropy and gluon density fluctuations within the nucleus.

Accurately determining the $t$ distribution also requires a thorough understanding of the transverse momentum distribution of the photons involved in photoproduction. Experimentally, this transverse momentum distribution cannot be directly measured; it is typically approximated using the Equivalent Photon Approximation (EPA)~\cite{Fermi:1924tc}, which inherently involves an integration over all impact parameters. However, recent theoretical and experimental studies of photon-photon collisions in heavy-ion collisions have demonstrated that the photon transverse momentum distribution highly depends on the collision impact parameter~\cite{Zha:2018tlq,CMS:2020skx,STAR:2019wlg,Klusek-Gawenda:2020eja}. This dependence necessitates a detailed investigation into the impact parameter dependence of exclusive photoproduction processes. In both electron-ion collisions and ultra-peripheral heavy-ion collisions (UPCs), traditional methods such as using charged-particle multiplicity to determine the impact parameter are not applicable. Recent measurements by the STAR~\cite{STAR:2019wlg}, ALICE~\cite{ALICE:2024ife,ALICE:2023jgu}, and CMS~\cite{CMS:2020skx,CMS:2023snh} experiments have successfully utilized neutron emission from Coulomb excitation of nuclei to effectively control the ``collision centrality'' in UPCs. This success motivates the application of a similar technique to control the impact parameter in electron-ion collisions, by tagging neutrons from Coulomb excitation to determine the interaction centrality.

To determine the probability of Coulomb dissociation (CD) as a function of impact parameter in UPCs, it is crucial to calculate the photon flux in spatial coordinates. In UPCs, the spatial distribution of the photon flux is typically computed using the EPA, which assumes a straight-line trajectory for the ions involved. This assumption holds well when the paths of the colliding ions can be considered unaffected by electromagnetic field over the collision duration. However, in electron-ion collisions at an EIC, the photon flux induced by the electron cannot be simply described by the conventional EPA, as the straight-line approximation breaks down due to significant deflection of the electron under the electromagnetic field of the heavy ion. This complication necessitates a precise derivation of the spatial distribution of the photon flux induced by the electron, which is essential for accurately calculating the CD probability versus impact parameter in electron-ion collisions.

This work aims to address these complexities by extending the traditional EPA framework to account for the unique dynamics of electron-induced photon flux in electron-ion collisions. By developing a spatially-dependent photon flux distribution, we aim to establish a more precise relationship between the photon transverse momentum distribution and impact parameter of collisions. Within this refined framework, we propose to study impact parameter manipulation for exclusive photoproduction processes in electron-ion collisions by tagging neutrons from Coulomb excitation. By achieving centrality control through neutron tagging, our study offers a new methodology for exploring the spatial and momentum structure of gluons in nuclei, ultimately contributing to the experimental design and data analysis strategies of future electron-ion collision experiments. 
%从Introduction到Methodology缺乏衔接
\section{Methodology}
\subsection{Kinematics of Electron-Proton/Nucleus Scattering}
\begin{figure}
    \centering
    \includegraphics[width=0.7\linewidth]{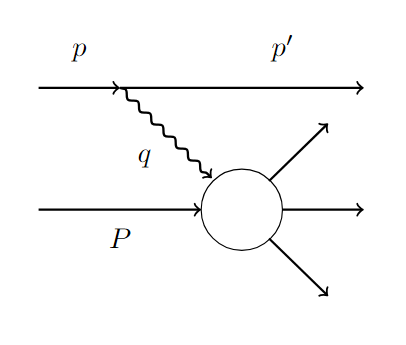}
    \caption{Feynman-like diagram for electron-proton scattering.}
    \label{epscattering}
\end{figure}

To derive the photon flux, we begin by analyzing the kinematics of electron-proton ($e + p$) scattering, as illustrated in Fig. \ref{epscattering}. Although our primary interest lies in electron-nucleus ($e + A$) collisions, the photon flux generated by the electron is essentially the same in both $e + p$ and $e + A$ interactions. Therefore, to keep the derivation straightforward while maintaining generality, we perform the analysis within the context of $e + p$ scattering.
The direction of the incident electron’s motion is chosen as the z-axis. The four vector of the incident electron $p$ and the scattering electron $p'$ are
\begin{equation}
p = \left(E_e, 0, 0, p_z\right)
\end{equation}
and
\begin{equation}
    p^\prime = \left(E_e', p_x, p_y, p_z'\right).
\end{equation}
The four-momentum $q$ of the emitted photon is
\begin{equation}
    q = \left(\omega, -p_x, -p_y, p_\gamma\right).
\end{equation}
The four vector conversion reads
\begin{align}
    p_z &= p_z' + p_\gamma, \\
    E_e &= E_e'+\omega, \\
    E_e^2 &= p_z^2 + m^2_e, \\
    p_T^2 = E_e'^2-p_z'^2-m^2_e &=\left(E_e-\omega\right)^2-\left(p_z-p_\gamma\right)^2-m^2_e, 
\end{align}
where $E_e$ and $E_e'$ are the energy of the incident electron and scattering electron, $m_e$ is the mass of the electron, $p_z$ and $p_z'$ are the z-component of the momentum of incident electron and scattering electron, $\omega$ is the energy of the virtual photon, and $p_T$ is the transverse momentum of the virtual photon. The momentum of the virtual photon is 
\begin{equation}
p_\gamma = p_z-\sqrt{\left(E_e-\omega\right)^2-p_T^2-m_e^2}.
\end{equation}
The virtuality of the photon can be written as
\begin{align}
    q^2 &= \omega^2-p_T^2-p_\gamma^2 \\
    &=\omega^2-p_T^2-\left(p_z-\sqrt{\left(E_e-\omega\right)^2-p_T^2-m_e^2}\right)^2.
\end{align}
The photon’s virtuality reaches its minimum and maximum values corresponding to two scenarios: the electron’s direction remains unchanged after scattering, and the electron’s direction is opposite to the original after scattering. Then the $q^2_{min}$ and $q^2_{max}$ can be written as
\begin{multline}
    q^2_{min} = 2E_e\omega-2E_e^2+2m^2_e +\\
    2\sqrt{\left(E_e^2-m^2_e\right)\left[\left(E_e-\omega\right)^2-m^2_e\right]}, \\
    q^2_{max} = \omega^2- \left[\sqrt{E_e^2-m^2_e}+\sqrt{\left(E_e-\omega\right)^2-m^2_e}\right]^2.
\end{multline}
The maximum value of the photon energy is $E_e-m$, naturally leads to
\begin{align}
q^2_{max}|_{\omega=E_e-m}=q^2_{min}|_{\omega=E_e-m}=&2m^2_e-2E_em, 
\end{align}
This indicates that the photon flux is zero at $\omega=\omega_{max}$, which is consistent with expectations. For $Q^2=-q^2\ll \omega^2$, the $Q_{min}^2$ and $Q_{max}^2$ can written as
\begin{align}
  Q_{min}^2&=\frac{m^2_e\omega^2}{E_e\left(E_e-\omega\right)} \label{q2old}, \\
  Q_{max}^2&=4E_e\left(E_e-\omega\right).   
\end{align}
\subsection{Photon Flux Derivation}
The cross section for process shown in Fig. \ref{epscattering} can be written as~\cite{Budnev:1975poe}
\begin{equation}
d \sigma_{e p}=\sigma_\gamma(\omega) d n,
\end{equation}
$\sigma_\gamma(\omega)$ is the absorption cross section for photons with frequency $\omega$, and $dn$ is the equivalent photon number. Representing the amplitude for virtual photon absorption as $M^{\mu}$, upon averaging over the initial spin states and summing across the final states, the cross section for electron-proton scattering is given by:
%\begin{widetext}
\begin{multline}
  \mathrm{d} \sigma_{\text {ep }} =\frac{4 \pi \alpha}{\left(-q^2\right)} M^{* \nu} M^\mu \rho^{\mu \nu} \\
  \times\frac{(2 \pi)^4 \delta\left(p+P-p^{\prime}-k\right) \mathrm{d} \Gamma}{4 \sqrt{(p P)^2-p^2 P^2}} \frac{\mathrm{d}^3 p^{\prime}}{2 E_e^{\prime}(2 \pi)^3},  
\end{multline}
where $\Gamma$ is the phase space volume, $\rho^{\mu \nu}$ is the density matrix of the virtual photon produced by an electron, written as:
\begin{align}
\rho^{\mu \nu} &=\frac{1}{2\left(-q^2\right)} \operatorname{Tr}\left[\left(\slashed{p}+m_e\right) \gamma^\mu\left(\slashed{p}^{\prime}+m_e\right) \gamma^\nu\right] \nonumber \\ 
&=-\left(g^{\mu \nu}-\frac{q^\mu q^\nu}{q^2}\right)-\frac{(2 p-q)^\mu(2 p-q)^\nu}{q^2}.
\label{photondensity}
\end{align}
For a nucleus with a defined charge distribution rather than a point-like particle, Eq. \ref{photondensity} can naturally be extended to:
\begin{align}
   \rho^{\mu \nu} &=\frac{1}{2\left(-q^2\right)} \operatorname{Tr}\left[\left(\slashed{p}+m\right) \gamma^\mu\left(\slashed{p}^{\prime}+m\right) \gamma^\nu\right] \nonumber\\
&=-\left(g^{\mu \nu}-\frac{q^\mu q^\nu}{q^2}\right)C\left(Q^2\right) \\
   &\quad-\frac{(2 p-q)^\mu(2 p-q)^\nu}{q^2}D\left(Q^2\right),
\label{photondensitywithform} 
\end{align}
where $C\left(Q^2\right)=F^2_M\left(Q^2\right)$ and $D\left(Q^2\right)=\frac{4m^2F_E^2+Q^2F_M^2}{4m^2+Q^2}$, $F^2_M$ and $F^2_E$ are the magnetic form factor and the electric form factor of the nucleus.
After integration over the phase space volume, the cross section can be written as
 \begin{multline}
\mathrm{d} \sigma=\frac{\alpha}{4 \pi^2\left|q^2\right|}\left[\frac{(q P)^2-q^2 P^2}{(p P)^2-p^2 P^2}\right]^{1 / 2} \times\\ \left(2 \rho^{++} \sigma_{\mathrm{T}}+\rho^{00} \sigma_{\mathrm{S}}\right) \frac{\mathrm{d}^3 p^{\prime}}{E_e^{\prime}},
\end{multline}
$\sigma_T$ and $\sigma_S$ are the cross section for transverse and scalar photon absorption respectively, $\sigma_S$ is negligible for quasi-real photon. The coefficients $\rho^{ab}$ are the elements of the density matrix in the $\gamma p$-helicity basis, written as:
\begin{equation}
2 \rho^{++}=\frac{(2 p P-q P)^2}{(q P)^2-q^2 P^2}+1+\frac{4 m_e^2}{q^2}, \  \rho^{00}=2 \rho^{++}-\frac{4 m_e^2}{q^2}-2.
\end{equation}
In the rest frame of the proton, i.e., the target frame, the following relationship holds true:
\begin{equation}
\omega=\frac{q P}{m_{\mathrm{p}}} , \quad E_e=\frac{p P}{m_{\mathrm{p}}} , \\
\quad \frac{\mathrm{d}^3 p^{\prime}}{E_e^{\prime}}=\frac{\mathrm{d} \omega \mathrm{d}\left(-q^2\right) \mathrm{d} \varphi}{2 \sqrt{E_e^2-m_{\mathrm{e}}^2}}. %\rightarrow \pi \frac{\mathrm{d} %\omega \mathrm{d}\left(-q^2\right)}{\sqrt{E_e^2-m_{\mathrm{e}}^2}},
\label{covariant}
\end{equation}
Let $Q^2=-q^2$, the equivalent photon number is 
%\begin{widetext}

\begin{align}
\frac{d^2n}{dQ^2 d\omega}&=\frac{\alpha}{2 \pi Q^2 E_e\left(E_e-m_e\right)} \rho^{++} \sqrt{\omega^2+Q^2}\nonumber \\
&=\frac{\alpha}{4 \pi Q^2 E_e\left(E_e-m_e\right)} \nonumber \\
&\times \left[\frac{(2 E_e-\omega)^2}{\omega^2+Q^2}+1-\frac{4 m_e^2}{Q^2}\right] \sqrt{\omega^2+Q^2}. 
\end{align}

\label{QEDphotoflux}
%\end{widetext}
The $\frac{d^2 n}{dQ^2d\omega}$ can be converted to $\frac{d^2n}{dp_T d\omega}$ by performing a variable change:
\begin{align}
dQ^2 d\omega &=
\begin{vmatrix}
\frac{\partial Q^2}{\partial p_T} & \frac{\partial Q^2}{\partial \omega} \nonumber \\
\frac{\partial \omega}{\partial p_T} & \frac{\partial \omega}{\partial \omega}
\end{vmatrix}
dp_Td\omega \\
&=\frac{2p_zp_T}{\sqrt{\left(E_e-\omega\right)^2-p_T^2-m_e^2}}dp_Td\omega,
\end{align}
   \begin{equation}
\frac{d^2n}{dp_Td\omega}=\frac{2p_zp_T}{\sqrt{\left(E_e-\omega\right)^2-p_T^2-m_e^2}}\frac{d^2n}{dQ^2d\omega}.
%\frac{\alpha}{4 \pi Q^2 E_e\left(E_e-m_e\right)}\left[\frac{(2 E_e-\omega)^2}{\omega^2+Q^2}+1-\frac{4 m_e^2}{Q^2}\right] \sqrt{\omega^2+Q^2}.
\label{dndpt}
\end{equation} 
The photon density matrix can be treated as the square of the photon wave function, thus the equivalent photon number in coordinate space can be obtained by performing a representation transformation on Eq. \ref{dndpt}:
\begin{equation}
\frac{d^3n}{d^2rd\omega}=\frac{\alpha}{\omega\pi^2}\left(\int_{0}^{p_{T_{max}}}\sqrt{\frac{p_T\pi\omega}{2\alpha}\frac{d^2n}{dp_Td\omega}}J_1\left(p_T\cdot r\right)\right)^2,
\label{photonflux_coordinate}
\end{equation}
$p_{T_{max}}$ is determined by $Q^2$, $E_e$ and $\omega$. Hereafter, we refer to the method of obtaining the photon flux in this manner as the QED approach.

For photoproduction in relativistic heavy-ion collisions, the photon flux is typically estimated using the classical EPA, which was independently derived by Williams~\cite{williams1933} and Weizsäcker~\cite{vonWeizsacker:1934nji} in the 1930s. In their derivation, they assumed that the charged particles move along straight-line trajectories, and obtained the spatial distribution of the electromagnetic field by solving the vector potential wave equation. The spatial distribution of the equivalent photon number was then derived based on the relationship between the energy flux density and the equivalent photon number. This approach provides an effective way to describe the photon flux distribution, which can be expressed as:
\begin{align}
    &n\left(\omega, \vec{x}_{\perp}\right)  =\frac{1}{\pi \omega}\left|\vec{E}_{\perp}\left(\omega, \vec{x}_{\perp}\right)\right|^2 \nonumber \\
 &=\frac{4 Z^2 \alpha}{\omega}\left|\int \frac{d^2 \vec{k}_{\perp}}{(2 \pi)^2} \vec{k}_{\perp} \frac{F\left(\vec{k}_{\perp}^2+\left(\frac{\omega}{\gamma}\right)^2\right)}{\vec{k}_{\perp}^2+\left(\frac{\omega}{\gamma}\right)^2} e^{i \vec{x}_{\perp} \cdot \vec{k}_{\perp}}\right|^2 \nonumber \\
 &=\frac{Z^2 \alpha}{\pi^2 \omega}\left|\int_0^{\infty} d k_{\perp} k_{\perp}^2 \frac{F\left({k_{\perp}}^2+\left(\frac{\omega}{\gamma}\right)^2\right)}{{k_{\perp}}^2+\left(\frac{\omega}{\gamma}\right)^2} J_1\left(x_{\perp} k_{\perp}\right)\right|^2,
\label{EPAphotonflux}
\end{align}
where Z is the charge number of the charged particle, $\gamma$ is the Lorentz factor of the charged particle, $\omega$ is the energy of the photon. For a point like particle, the photon flux is
\begin{equation}
n_{\mathrm{pt}}\left(\omega, x_{\perp}\right)=\frac{Z^2 \alpha_{Q E D} \omega}{\pi^2 \gamma^2}\left[K_1\left(\frac{\omega x_{\perp}}{\gamma}\right)\right]^2.
\end{equation}

\subsection{Coulomb Dissociation in Electron-Ion Collisions}

Analogous to the Coulomb excitation process in relativistic heavy-ion collisions, Coulomb excitation in electron-ion collisions can be factorized into two distinct components: the emission of quasi-real photons by the electron, and the corresponding photon absorption cross section of the nucleus. The quasi-real photons emitted by the electron can be estimated using the framework described in the previous subsection. 

The lowest-order probability for a nucleus to be excited to a state that subsequently emits at least one neutron (denoted as Xn) can be expressed as~\cite{Baltz:1996as}:
\begin{equation}
    m_{Xn}(b) = \int d \omega \, n(\omega, b) \, \sigma_{Xn, \gamma A \rightarrow A^*}(\omega),
\end{equation}
where $n(\omega, b)$ represents the photon flux at a given impact parameter $b$ and $\sigma_{Xn, \gamma A \rightarrow A^*}(\omega)$ is the photoexcitation cross section for an incident photon with energy $\omega$, obtained from experimental data~\cite{Veyssiere:1970ztg, Lepretre:1981tf, Carlos:1984lvc, Caldwell:1973bu, Armstrong:1972sa}.

It is important to note that, under certain conditions—such as very small impact parameters and extremely high beam energies—the value of $m_{Xn}$ could exceed 1, implying that the excitation probability would lose its probabilistic interpretation. Although such conditions are unlikely to occur at current or near-future facilities, it is useful to address this scenario for the sake of completeness. To maintain a valid probabilistic interpretation, $m_{Xn}(b)$ is treated as the mean number of photons absorbed by the nucleus, and we assume that the photon multiplicity follows a Poisson distribution~\cite{Pshenichnov:2001qd,Pshenichnov:1999hw}. In this context, the probability for absorbing zero photons (i.e., zero neutron emission) is given by:
\begin{equation}
    P^{(0)}(b) = e^{-m_{Xn}(b)},
\end{equation}
while the probability of absorbing exactly $N$ photons is
\begin{equation}
    P^{(N)}(b) = \frac{m_{Xn}^{N}(b)}{N!} e^{-m_{Xn}(b)}.
\end{equation}

The normalized probability density for the absorption of one photon with energy $E_1$ can be expressed as:
\begin{equation}
    p^{(1)}(E_1, b) = \frac{n(E_1, b) \, \sigma_{\gamma A \rightarrow A^*(E_1)}}{m_{Xn}(b)},
\end{equation}
and the probability density for absorbing $N$ photons with energies $E_1, E_2, \ldots, E_N$ is:
\begin{equation}
    p^{(N)}(E_1, E_2, \ldots, E_N, b) = \frac{\prod_{i=1}^N n(E_i, b) \, \sigma_{\gamma A \rightarrow A^*(E_i)}}{m_{Xn}(b)}.
\end{equation}

For a specific electromagnetic dissociation channel involving the emission of $i$ neutrons, the probability density for an $N$-photon absorption process can be evaluated as:
\begin{multline}
    P_i^{(N)}(b) = \int \ldots \int dE_1 \ldots dE_N \\
    \times P^{(N)}(b) \, p^{(N)}(E_1, \ldots, E_N, b) \, f_i(E_1, \ldots, E_N),
\end{multline}
where $f_i(E_1, \ldots, E_N)$ represents the branching ratio for the specific channel with $i$ neutrons emitted. We assume that the simultaneous absorption of multiple photons is allowed, leading to a simplified form for the branching ratio: $f_i(E_1, \ldots, E_N) = f_i\left(\sum_{k=1}^{N} E_k\right)$. The values of $f_i$ for different neutron emission channels are extracted from the $n_O^On$ model as detailed in Ref.~\cite{Broz:2019kpl}.

Finally, the total probability for the emission of $i$ neutrons is given by:
\begin{equation}
    P_{in}(b) = \sum_{k=1}^\infty P_i^{(k)}(b).
\end{equation}
\begin{figure*}
    \centering
    \subfigure[Photon flux from EPA]{\includegraphics[width=0.3\linewidth]{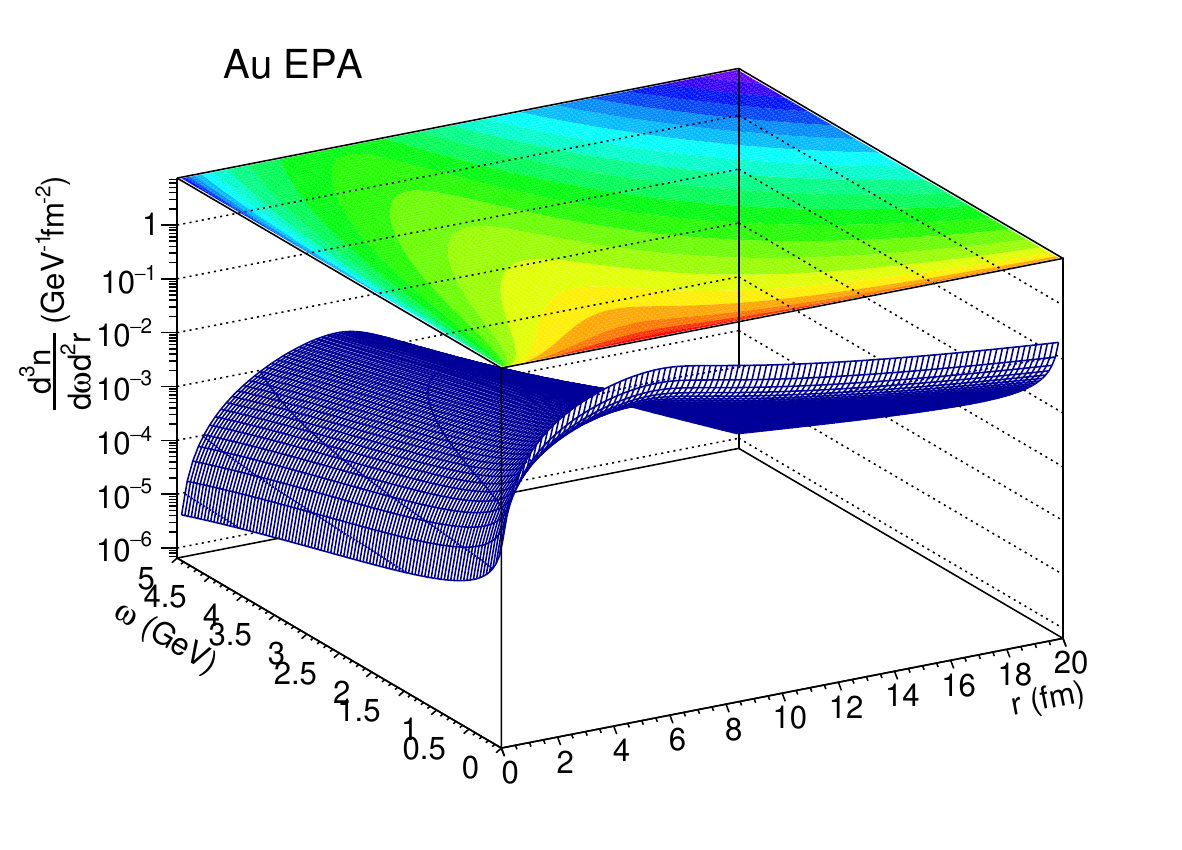}}
    \subfigure[Photon flux from QED]{\includegraphics[width=0.3\linewidth]{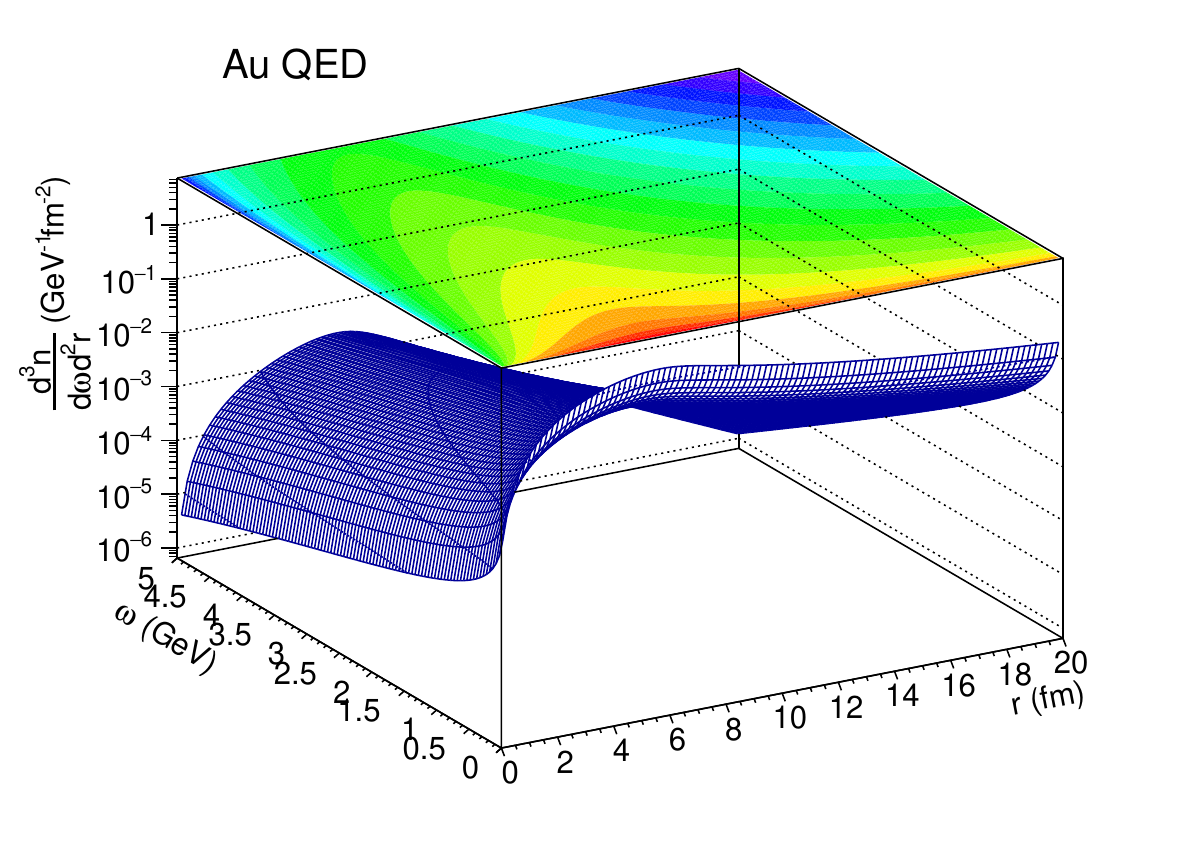}}
    \subfigure[QED to EPA photon flux ratio]{\includegraphics[width=0.3\linewidth]{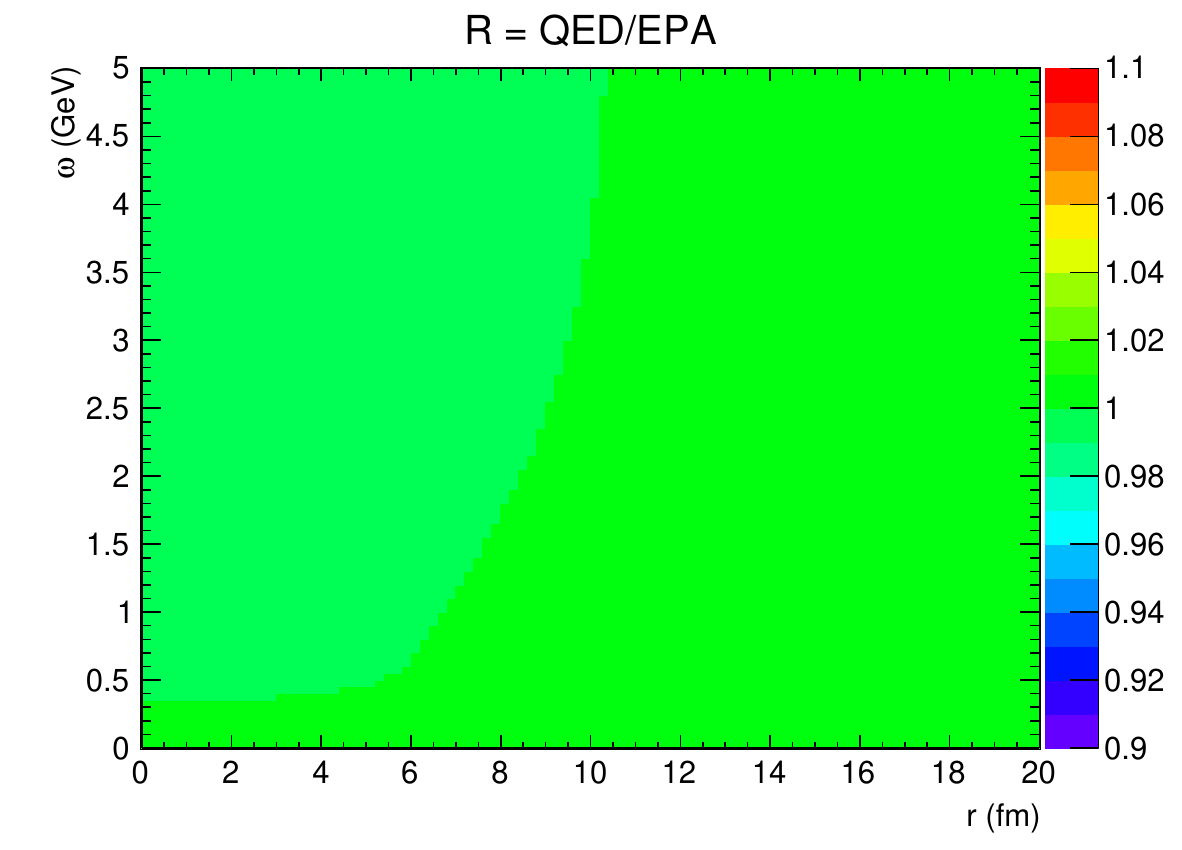}}
    \caption{Comparison of the photon flux distribution induced by an Au nucleus with $E$ = 100 GeV per nucleon, as calculated using the EPA model and the QED derivation, along with the ratio of the QED results to the EPA results.}
    \label{Auphotonflux}
\end{figure*}

\begin{figure*}
    \centering
    \subfigure[Photon flux from EPA]{\includegraphics[width=0.3\linewidth]{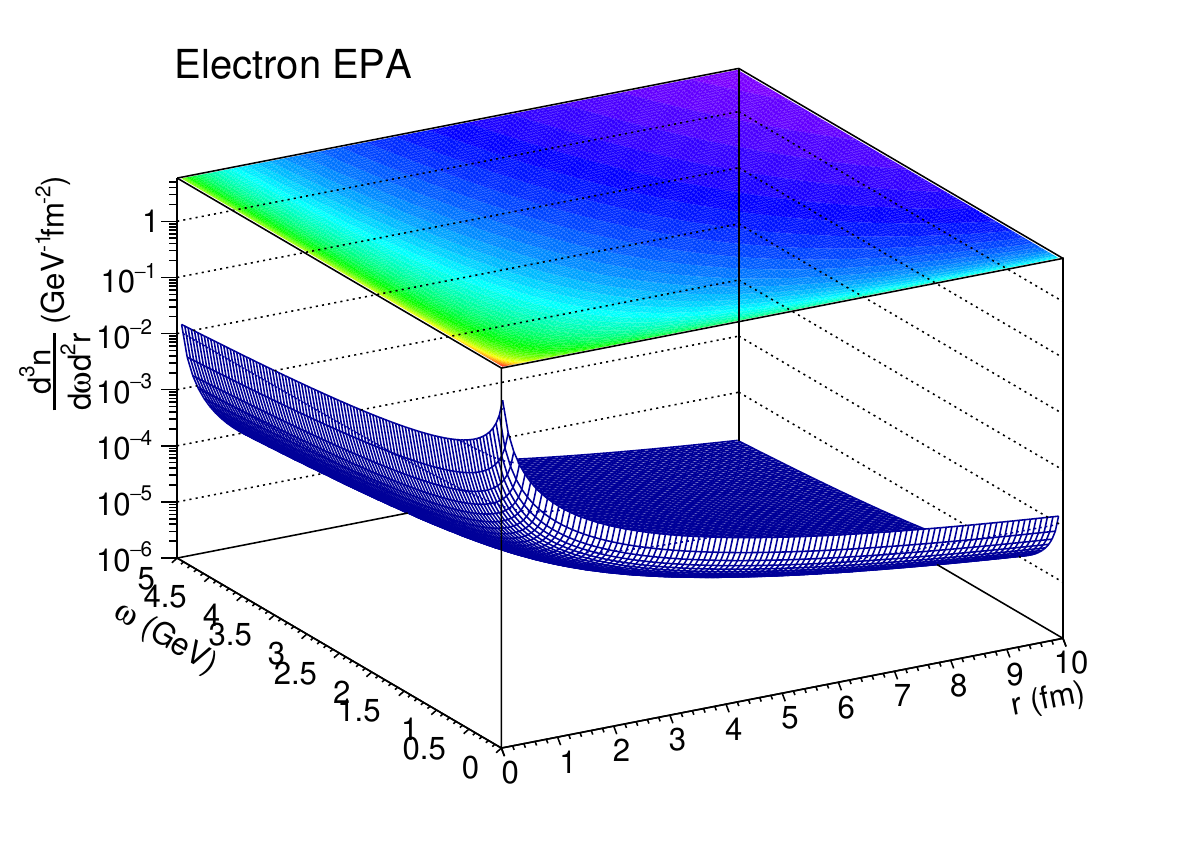}}
    \subfigure[Photon flux from QED]{\includegraphics[width=0.3\linewidth]{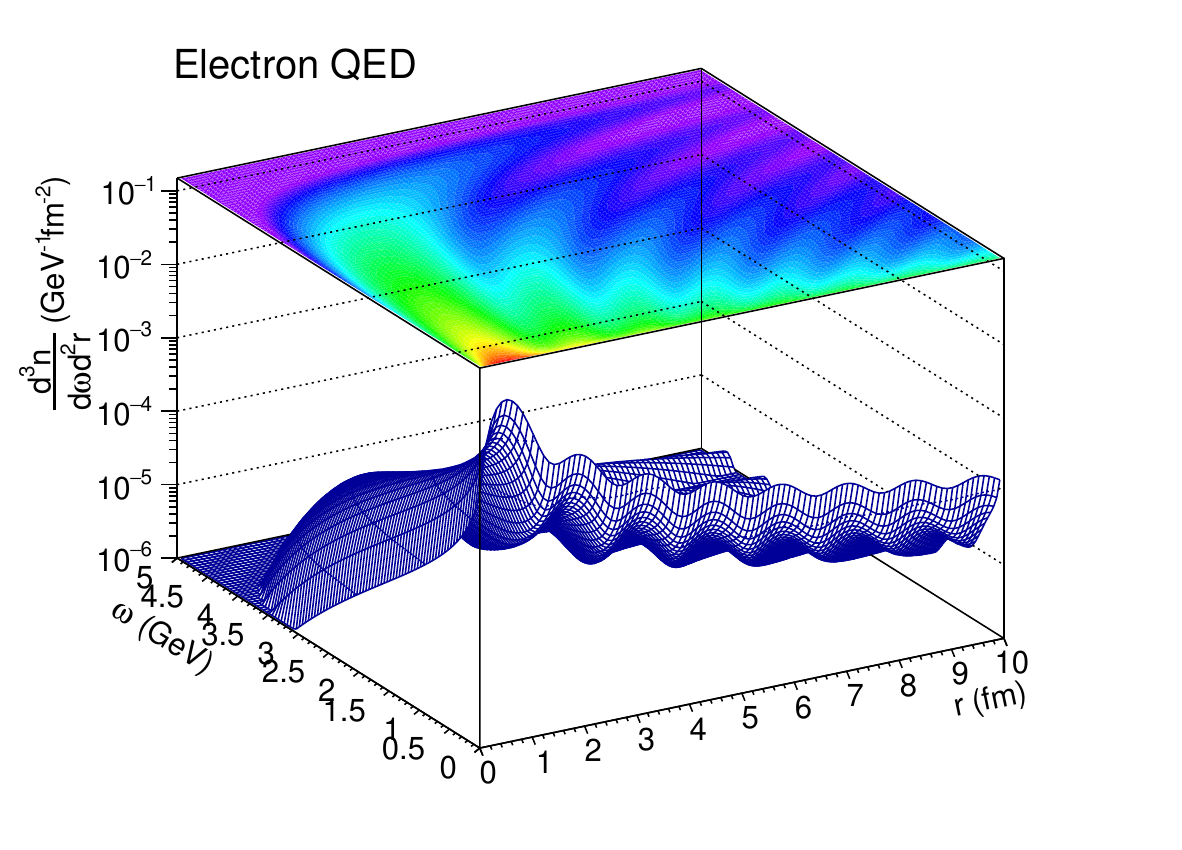}}
    \subfigure[QED to EPA photon flux ratio]{\includegraphics[width=0.3\linewidth]{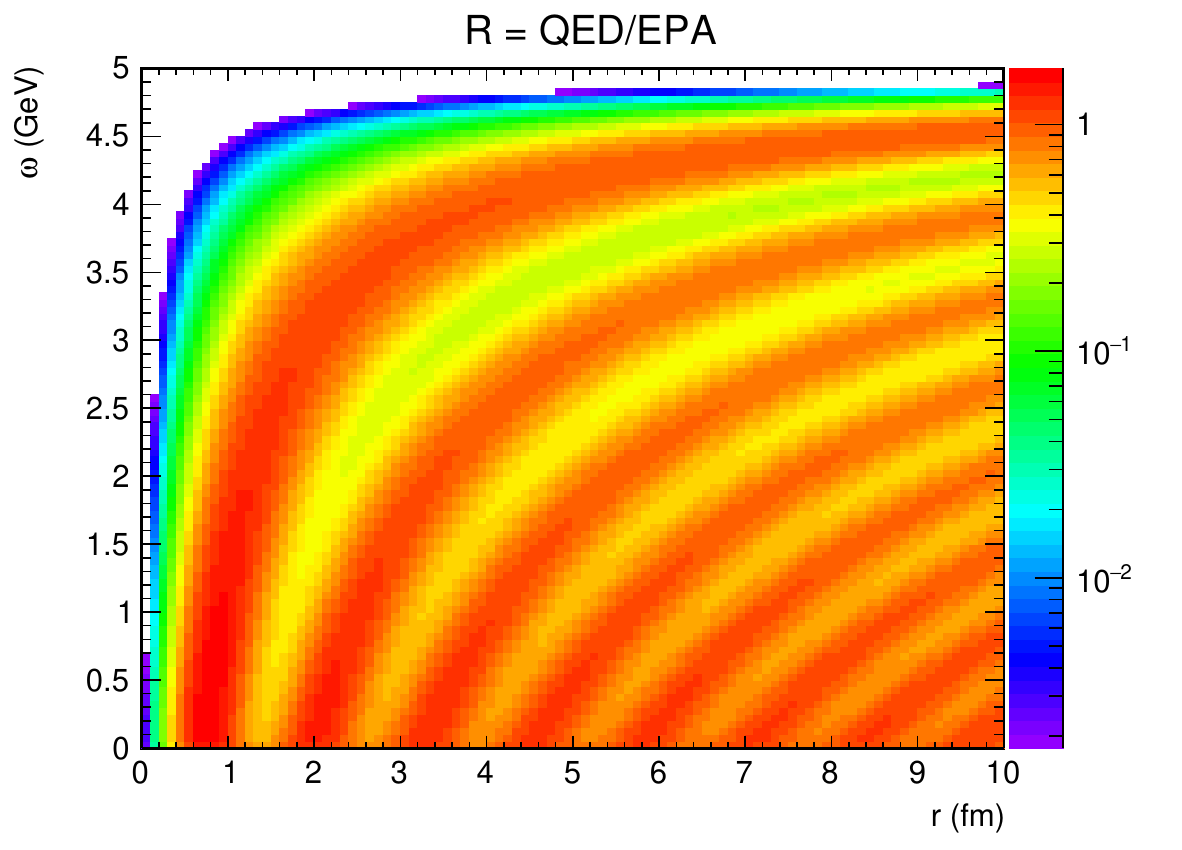}}
    \caption{Comparison of the photon flux distribution induced by an electron with $E$ = 5 GeV, as calculated using the EPA model and the QED derivation, along with the ratio of the QED results to the EPA results.}
    \label{electronphotonflux}
\end{figure*}

\subsection{Vector Meson Photoproduction in Electron-Ion Collisions}

The estimation of vector meson photoproduction in electron-ion collisions can be performed in a manner similar to the Coulomb excitation calculations. The main difference lies in replacing the photon absorption cross section of the nucleus with the $\gamma A \rightarrow V A$ cross section. Specifically, the scattering amplitude $\Gamma_{{\gamma}A \rightarrow V A}$, including the shadowing effect, can be derived using the Glauber model~\cite{Miller:2007ri} combined with the vector meson dominance (VMD) approach~\cite{Bauer:1977iq}:

\begin{multline}
    \Gamma_{{\gamma}A \rightarrow V A}(\vec{x}_\perp) = \frac{f_{{\gamma}N \rightarrow V N}(0)}{\sigma_{VN}} \times \\
    2 \left[ 1 - \mathrm{exp}\left( -\frac{\sigma_{VN}}{2} T^{\prime}\left(\vec{x}_\perp\right) \right) \right], \label{5}
\end{multline}
where $f_{{\gamma}N \rightarrow V N}(0)$ is the forward-scattering amplitude for $\gamma + N \rightarrow V + N$, and $\sigma_{VN}$ represents the total vector meson-nucleon ($V N$) cross section. The modified nuclear thickness function $T^{\prime}(\vec{x}_\perp)$, which takes into account the coherence length effect, is given by:

\begin{equation}
    T^{\prime}(\vec{x}_\perp) = \int_{-\infty}^{+\infty} \rho\left(\sqrt{{\vec{x}_\perp}^2 + z^2}\right) e^{iq_L z} \, dz, \quad q_L = \frac{M_V e^y}{2 \gamma_c}, \label{6}
\end{equation}
where $q_L$ is the longitudinal momentum transfer required to produce a real vector meson, $M_V$ is the vector meson mass, and $\gamma_c$ is the Lorentz factor of the nucleus.

Considering the impact of the photon's virtuality on the photon-nucleon scattering cross section, the equivalent vector meson flux is introduced:
\begin{equation}
    \frac{d^2V}{d\omega dQ^2} = \left( \frac{M_V^2}{M_V^2 + Q^2} \right)^n \frac{d^2n}{d\omega dQ^2},
\end{equation}
where $\left( \frac{M_V^2}{M_V^2 + Q^2} \right)^n$ represents the suppression factor associated with the transition amplitude from the virtual photon fluctuation to the corresponding vector meson, and $n$ is determined by fitting experimental data~\cite{H1:1999pji,H1:2009cml}. The equivalent vector meson flux in coordinate space, $\frac{d^3n}{d^2r d\omega}$, can be obtained using the method outlined in the previous section.

The amplitude distribution for the vector meson photoproduction process is given by:
\begin{equation}
    A\left(b, \vec{x}_\perp\right) = \Gamma_{{\gamma}A \rightarrow V A}\left(\vec{r}_{1}\right) \sqrt{n\left(\omega, \vec{r}_2\right)},
\end{equation}
where $\vec{r}_{2} - \vec{r}_{1} = \vec{b}$ and $\frac{\vec{b}}{2} + \vec{r}_{1} = \vec{x}_\perp$. The production amplitude in momentum space can be obtained by applying a Fourier transformation to the amplitude in coordinate representation:
\begin{equation}
    \vec{A}(\vec{p}_\perp, b) = \frac{1}{2\pi} \int d^2x_\perp \, \vec{A}(\vec{x}_\perp, b) e^{i\vec{p}_\perp \cdot \vec{x}_\perp}. \label{APT}
\end{equation}
From Eq.~\ref{APT}, the differential cross section $\frac{d\sigma}{dt}$ can be calculated.

Finally, the photoproduction cross section in conjunction with the Coulomb excitation of the nucleus can be estimated as follows:
\begin{multline}
    \frac{d\sigma_{e A \rightarrow e A^* + Xn}}{dY} = \int d^2 \vec{x}_\perp \int \omega \left| A\left(b, \vec{x}_\perp\right) \right|^2 \\
    \times P_{in}(b) \, 2\pi b \, db,
\end{multline}
where $Y$ is the rapidity of the photoproduced vector meson, and $\omega = \frac{1}{2} M_{V} e^Y$. Here, $P_{in}(b)$ represents the probability of emitting $i$ neutrons, which accounts for the Coulomb excitation contribution.

\section{RESULTS}

\begin{figure}
    \centering
    \includegraphics[width=1.0\linewidth]{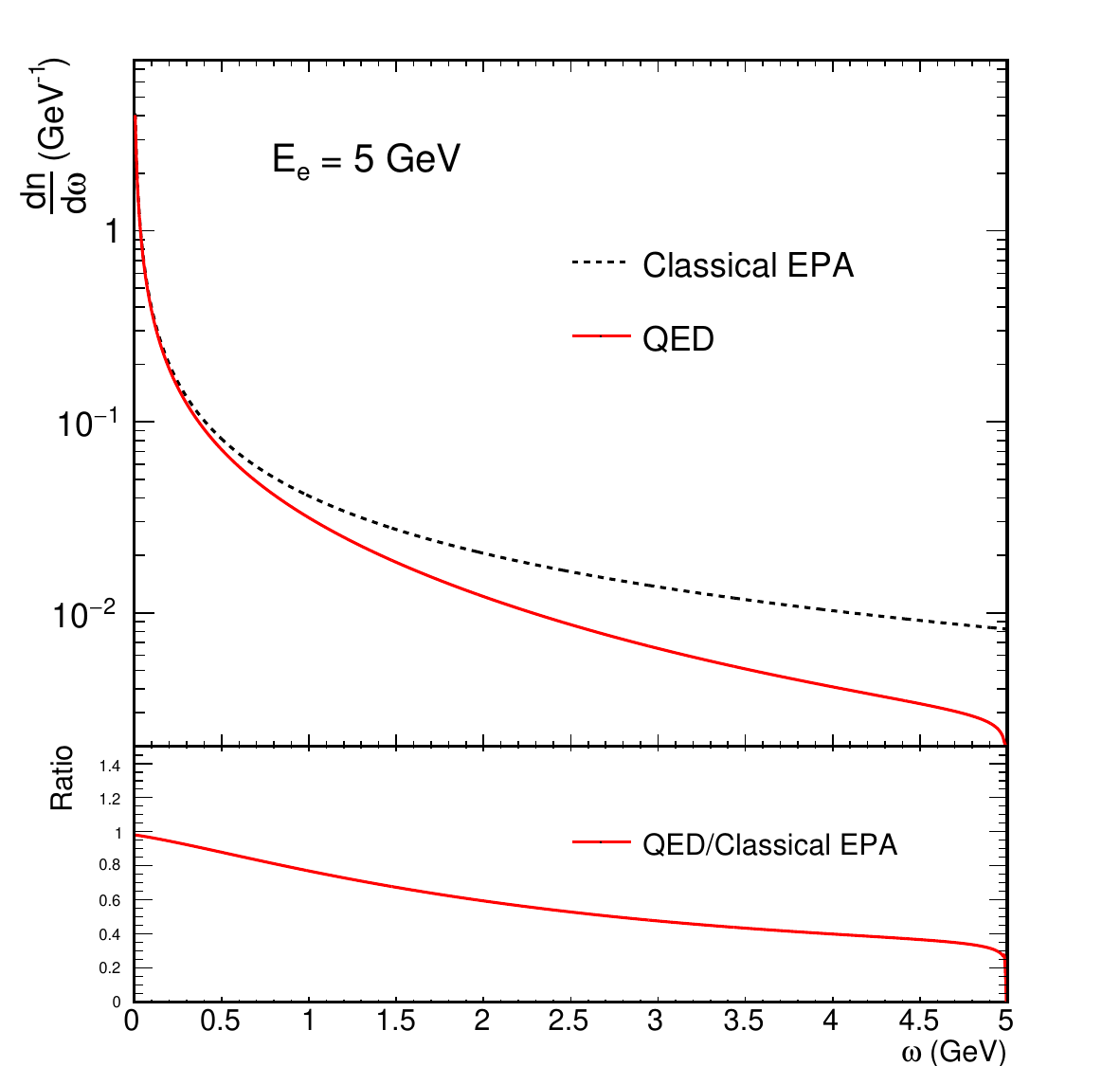}
    \caption{The $\frac{dn}{d\omega}$ distribution calculated from the classical EPA model and the QED model.}
    \label{photonflux_ratio_omega}
\end{figure}

\begin{figure}
    \centering
    \includegraphics[width=1.0\linewidth]{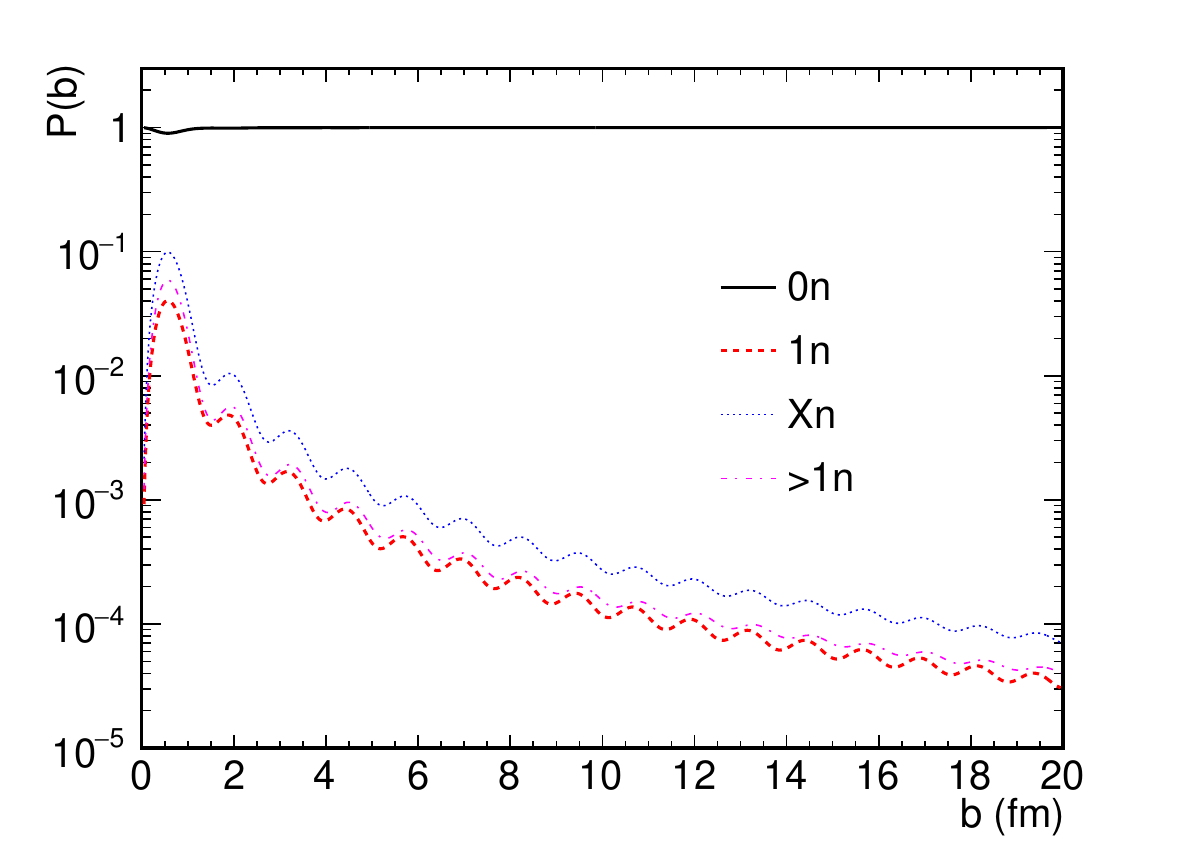}
    \caption{Nucleus break-up probability of Au-197 as a function of impact parameter in e+Au collision at EIC energy (18$\times$100 GeV) for different number of neutron emission.}
    \label{prob_EM_Xn}
\end{figure}

\begin{figure*}
    \centering
    \subfigure[]{\includegraphics[width=0.45\linewidth]{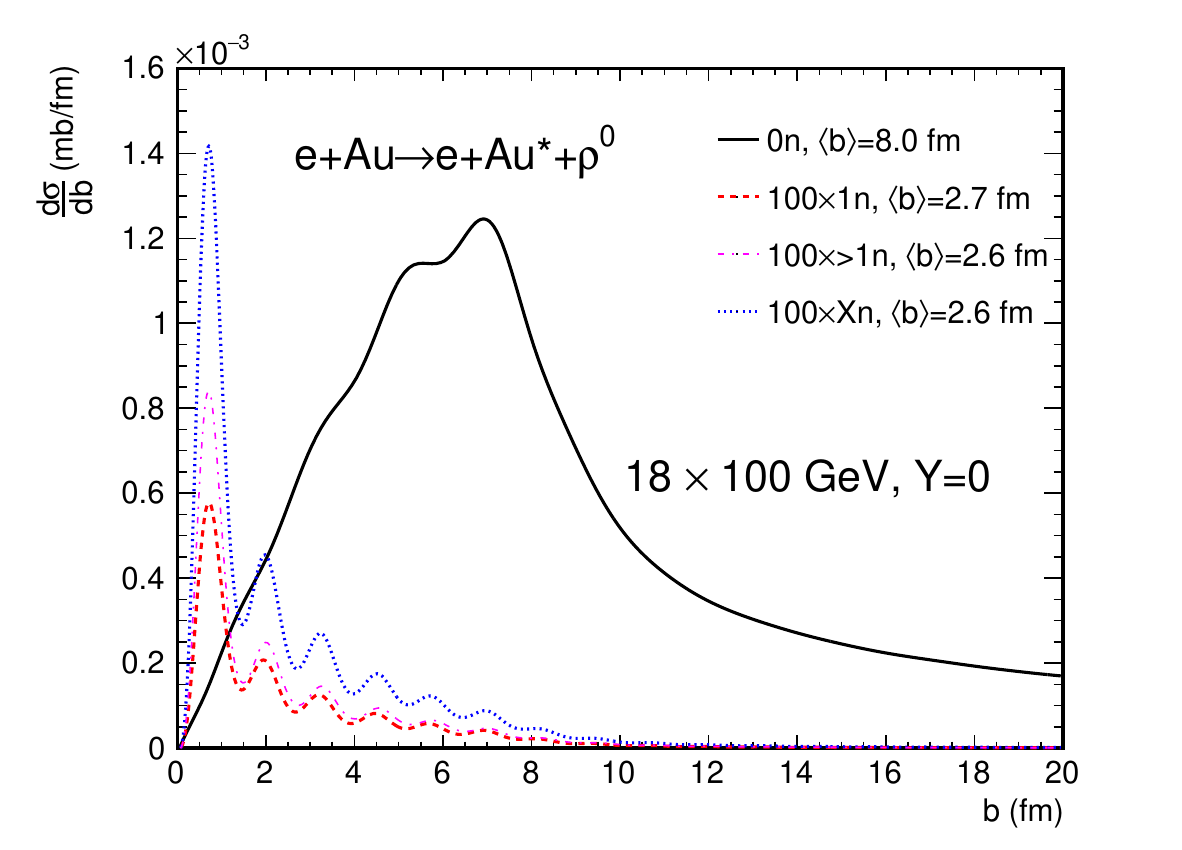}}
    \subfigure[]{\includegraphics[width=0.45\linewidth]{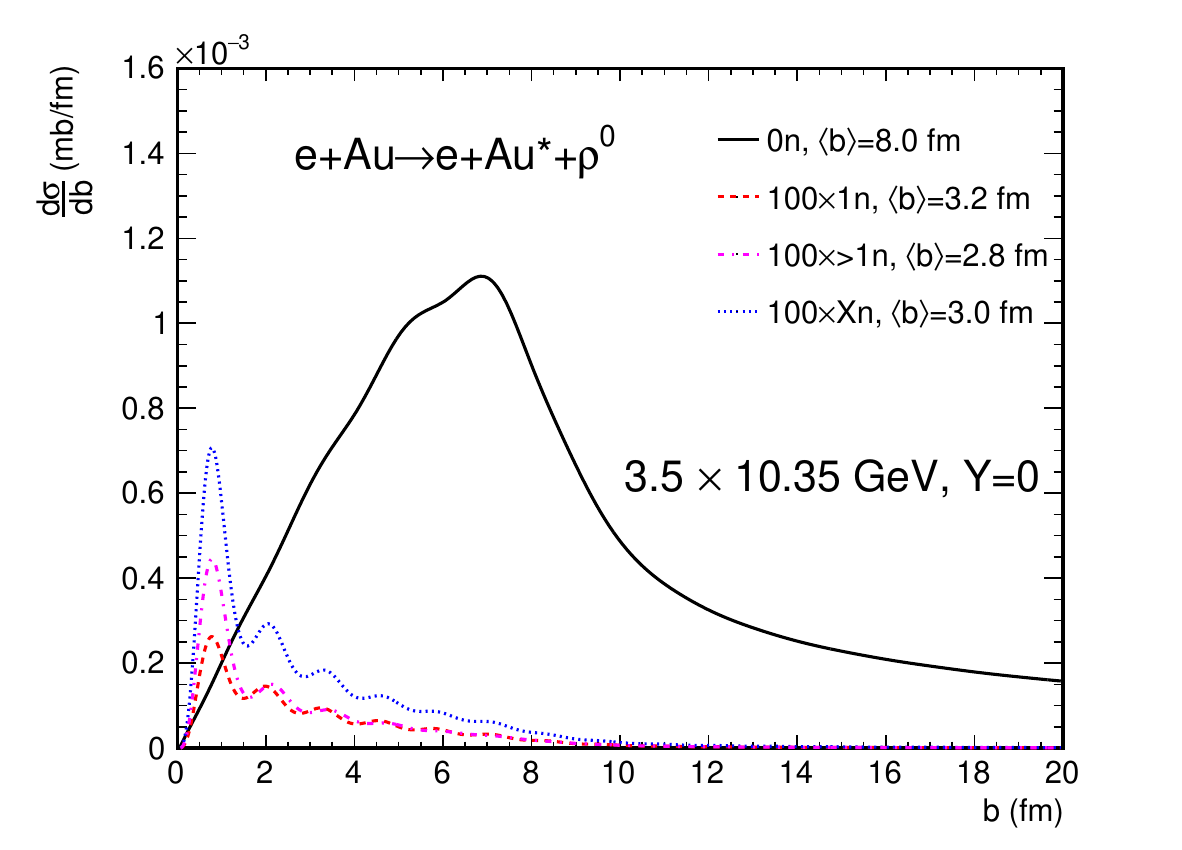}}
    \caption{The $\frac{d\sigma}{db}$ for $\rho^0$ photoproduction at EIC (a) and EicC (b) energy. Black line: ``0n'' mode. Blue line: ``1n'' mode. Red line: ``Xn'' mode. The results of ``1n'', ``>1n'' and ``Xn'' have been multiplied by 100.}
    \label{rhobdistribution}
\end{figure*}

\begin{figure*}
    \centering
    \subfigure[]{\includegraphics[width=0.45\linewidth]{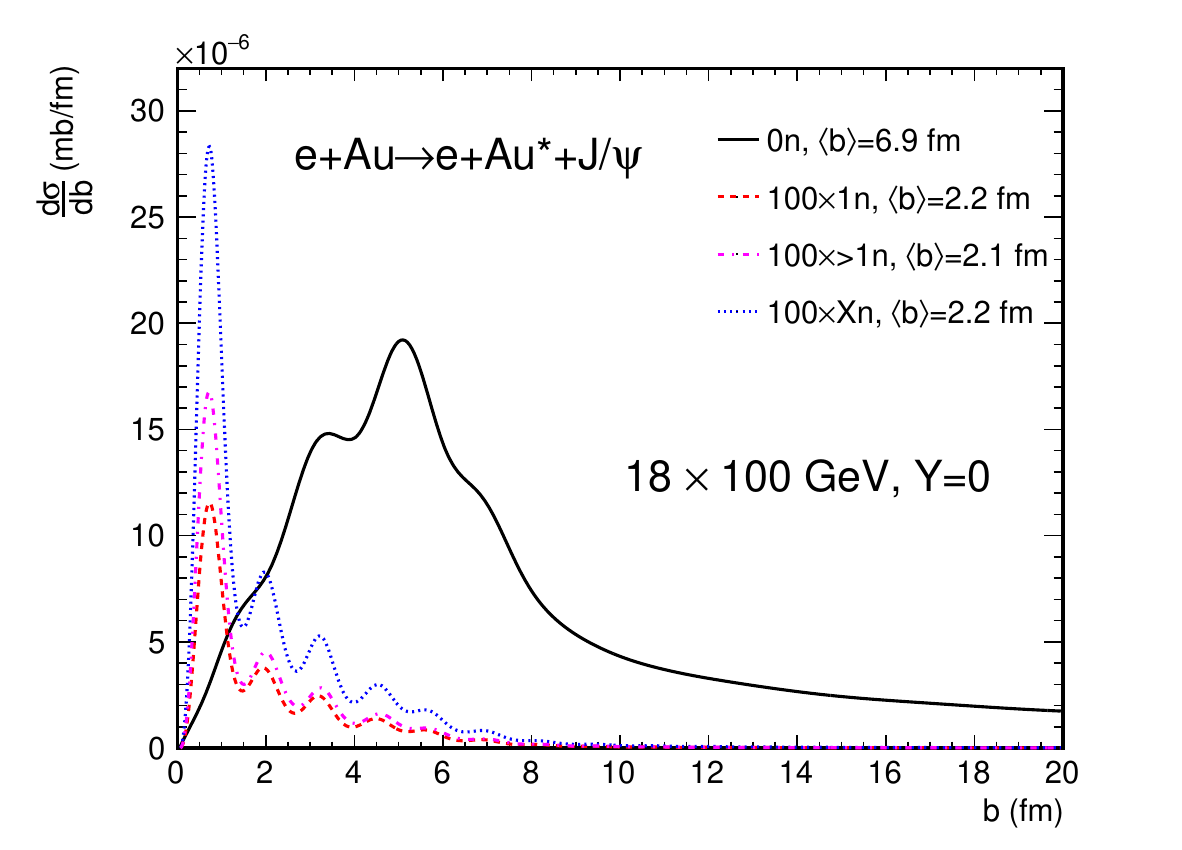}}
    \subfigure[]{\includegraphics[width=0.45\linewidth]{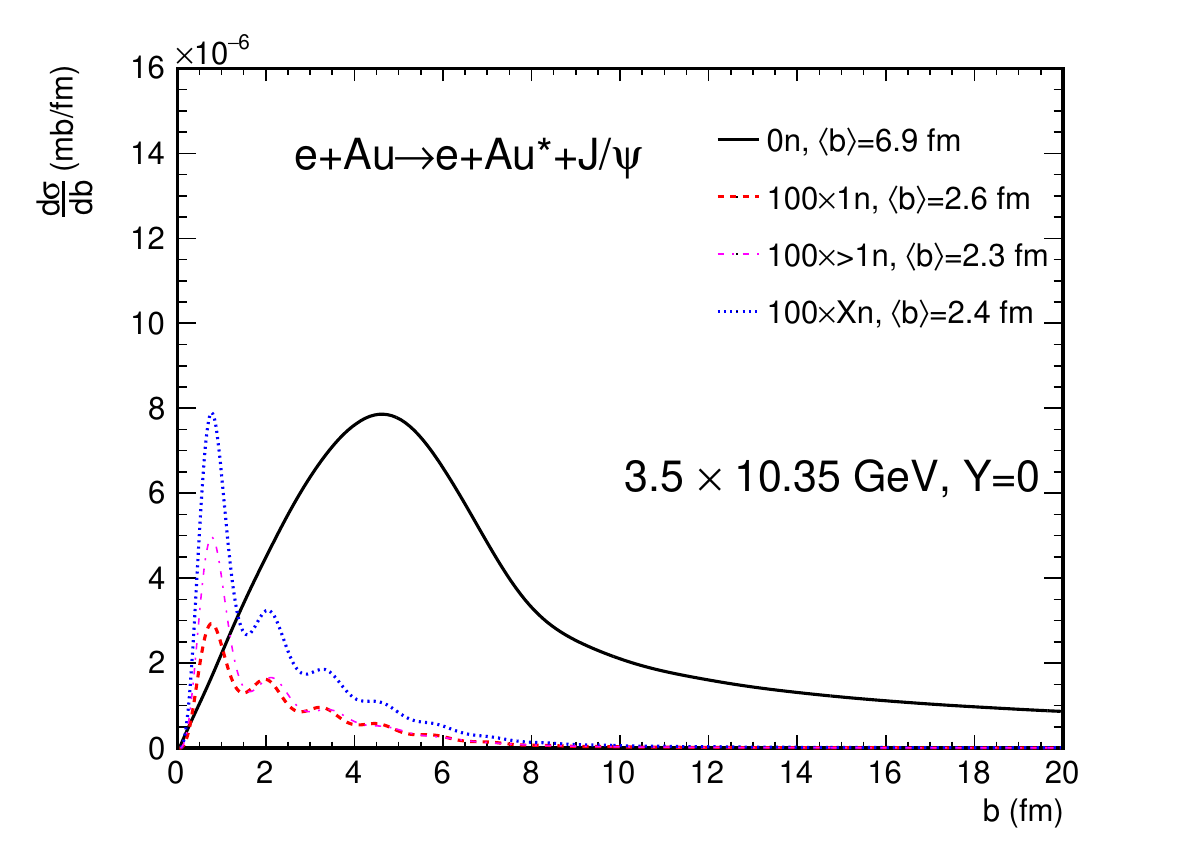}}
    \caption{The $\frac{d\sigma}{db}$ for $J/\psi$ photoproduction at EIC (a) and EicC (b) energy. Black line: ``0n'' mode. Blue line: ``1n'' mode. Red line: ``Xn'' mode. The results of ``1n'', ``>1n'' and ``Xn'' have been multiplied by 100.}
    \label{Jpsibdistribution}
\end{figure*}

In ultra-peripheral heavy-ion collisions, the photon flux is typically calculated using the EPA approach, as given by Eq.~\ref{EPAphotonflux}. This model is based on the assumption that the charged particles involved move along straight-line trajectories. However, questions have arisen regarding the validity of this assumption, particularly at RHIC and LHC energies. To examine the applicability of the EPA model, we compare it with the QED approach, which does not rely on the straight-line assumption. Figure \ref{Auphotonflux} presents the photon flux distribution induced by an Au nucleus with an energy of 100 GeV per nucleon, calculated using both the classical EPA and the QED models, along with the ratio of the QED results to the EPA results. From the figure, it can be observed that both models predict that the photon flux reaches its maximum value at the radius of the Au nucleus and decreases as the photon energy $\omega$ increases. Moreover, the ratio between the QED and EPA results remains close to unity, suggesting that the EPA model provides an accurate approximation of the photon flux for UPCs and is effectively equivalent to the QED-derived expression under these conditions.

However, in the case of electron-ion collisions, the use of Eq.~\ref{EPAphotonflux} becomes problematic. This is primarily because the energy of the electron is significantly smaller compared to that of the heavy ion, rendering the straight-line approximation invalid. Furthermore, a direct application of Eq.~\ref{EPAphotonflux} does not prohibit the photon energy from exceeding the energy of the charged particle itself, which is physically incorrect. To illustrate this limitation, we compare the photon flux distribution calculated using the classical EPA and the QED models for an electron with an energy of 5 GeV. Figure \ref{electronphotonflux} shows the 2D photon flux distribution and the ratio of the QED to EPA results. It is evident that the results obtained from the two models differ significantly, with the QED-derived flux showing distinct fluctuations and tending towards zero as the photon energy approaches the electron energy. This behavior highlights the inadequacy of the EPA model in describing the photon flux for electron-ion collisions.

Figure \ref{photonflux_ratio_omega} provides a further comparison of the photon energy distributions obtained from the EPA and QED models. It can be seen that the photon flux calculated using the QED model matches the EPA prediction at low photon energies but rapidly approaches zero as the photon energy nears the electron's total energy. In contrast, the photon flux calculated using the EPA model continues to decrease smoothly without reaching zero. This discrepancy further underscores the limitations of the EPA model for electron-ion collisions and demonstrates that the photon flux distribution derived from the QED model is more suitable for accurately describing these processes. Consequently, the QED approach provides a more reliable basis for calculating the impact parameter dependence of photoproduction processes in electron-ion collisions.

With the QED-derived photon flux, we can accurately evaluate the Coulomb excitation of a nucleus in electron-ion collisions. As an illustrative example, we consider $e + Au$ collisions at EIC energies, specifically at $18 \times 100$ GeV per nucleon. The corresponding $P_{in}(b)$ distribution, which represents the dissociation probability as a function of impact parameter, is shown in Fig. \ref{prob_EM_Xn}. The dissociation probability, characterized by neutron emission, decreases rapidly with increasing impact parameter. Interestingly, the probability distribution exhibits an oscillatory pattern, which can be attributed to the wave nature of the photons emitted by the electron, resulting in interference effects. Dissociation processes involving a higher number of emitted neutrons are more likely to occur at smaller impact parameters, indicating stronger electromagnetic interactions in more central collisions. This characteristic offers a practical method for determining the centrality in electron-ion interactions by counting the number of emitted neutrons detected by the Zero Degree Calorimeter (ZDC). The correlation between neutron multiplicity and impact parameter provides an effective means to categorize collision events by their geometric overlap, enabling a more precise study of photonuclear interaction dynamics in electron-ion collisions.

It is also important to note that different photoproduction processes, even with the same neutron tagging, will exhibit variations in impact parameter distributions. This is because the energy of the photon involved in different processes varies, which in turn affects the photon's spatial distribution relative to the electron. Consequently, unlike hadronic heavy-ion collisions, where the centrality is associated with a fixed impact parameter range, centrality determination in electron-ion collisions via neutron tagging depends on the specific photoproduction process under consideration and must be evaluated on a case-by-case basis. To illustrate this, we present calculations for coherent $\rho^0$ and $J/\psi$ photoproduction accompanied by different neutron tagging at EIC and EicC energies.

Figure \ref{rhobdistribution} presents the $d\sigma/db$ distributions for coherent $\rho^0$ photoproduction at EIC and EicC energies, with different line types and colors representing various neutron emission modes. The average impact parameter for the $0n$ mode is significantly larger than that for the $1n$ and $Xn$ (at least one neutron) modes. This is because neutron excitation requires additional photons, which consequently reduces the average impact parameter. This distinct variation in impact parameters across neutron emission modes demonstrates the feasibility of determining the centrality of electron-ion collisions by tagging neutrons from Coulomb excitation. Moreover, the cross section exhibits fluctuations with respect to the impact parameter, a phenomenon that arises from the oscillatory behavior of the $J_1$ Bessel function in the coordinate distribution of the photon flux, as described by Eq. \ref{photonflux_coordinate}. The comparison between the results at EIC and EicC energies indicates that varying the center-of-mass collision energy has a negligible effect on the average impact parameter, suggesting that the proposed method is effective across different collision energy regimes.

Figure \ref{Jpsibdistribution} illustrates the $d\sigma/db$ distributions for coherent $J/\psi$ photoproduction at EIC and EicC energies. Similar to the $\rho^0$ case, the average impact parameter for the $0n$ mode is much larger than that of other neutron emission modes. Additionally, the average impact parameter $\left<b\right>$ for $J/\psi$ is generally smaller than that for $\rho^0$ in the corresponding neutron emission modes. This behavior can be attributed to the larger mass of $J/\psi$ compared to $\rho^0$, which corresponds to a higher photon energy. Consequently, the photon is positioned closer to the electron, resulting in a reduced average impact parameter. The sensitive dependence of the average impact parameter on the number of neutrons emitted via Coulomb excitation, observed across different vector mesons, further underscores the effectiveness of this method for determining centrality in experimental electron-ion collisions.

\section{summary}
In summary, we have investigated the feasibility of employing neutron tagging resulting from the Coulomb excitation of nuclei as a means to precisely ascertain the centrality of exclusive photoproduction events in electron-ion collisions. By developing an equivalent photon approximation for electrons, this paper integrates a photon flux distribution in coordinate space, clarifying the connection between the distribution of the photon's transverse momentum and the impact parameters of the collisions. By leveraging the spatial data of the photon flux, the differential cross section for the Coulomb excitation of nuclei is calculated. Our calculations suggest that neutron tagging can markedly shift the distributions of impact parameters, thus offering a reliable technique for controlling the centrality in electron-ion collision experiments. This research provides essential references and approaches for examining the dependence of exclusive photoproduction processes on impact parameters, yielding novel perspectives for the design of experiments and the analysis of data.

\end{document}